\documentclass[fleqn,usenatbib,useAMS]{mnras}

\usepackage{newtxtext,newtxmath}
\usepackage[T1]{fontenc}
\usepackage{ae,aecompl}

\usepackage{graphicx}	% Including figure files

\usepackage[normalem]{ulem}
\newcommand{\eq}[1]{\begin{equation} #1
\end{equation}}

\newcommand\lsim{\mathrel{\rlap{\lower4pt\hbox{\hskip1pt$\sim$}}
        \raise1pt\hbox{$<$}}}
\newcommand\gsim{\mathrel{\rlap{\lower4pt\hbox{\hskip1pt$\sim$}}
        \raise1pt\hbox{$>$}}}

\title[Constraints on WDM from UV LFs]{Constraints on warm dark matter from UV luminosity functions of high-$z$ galaxies with Bayesian model comparison}

\author[A. Rudakovskyi et al.]{
Anton Rudakovskyi,$^{1,2}$\thanks{E-mail: rudakovskyi@bitp.kiev.ua}
Andrei Mesinger,$^{3}$
Denys Savchenko$^{1,2}$
and Nicolas Gillet$^{4}$
\\
$^{1}$Bogolyubov Institute for Theoretical Physics of the NAS of Ukraine, Metrolohichna Str. 14-b, Kyiv, 03143, Ukraine\\
$^{2}$Kyiv Academic University, 36 Vernadsky blvd., Kyiv, 03142, Ukraine\\
$^{3}$Scuola Normale Superiore, Piazza dei Cavalieri 7, I-56126 Pisa, Italy\\
$^{4}$Observatoire Astronomique de Strasbourg, Université de Strasbourg, CNRS UMR 7550, 11 rue de l’Université, 67000 Strasbourg, France
}

\date{Accepted XXX. Received YYY; in original form ZZZ}

\pubyear{2021}

\begin{document}
\label{firstpage}
\pagerange{\pageref{firstpage}--\pageref{lastpage}}
\maketitle

% Abstract of the paper
\begin{abstract}
The number density of small dark matter (DM) halos hosting faint high-redshift galaxies is sensitive to the DM free-streaming properties. However, constraining these DM properties is complicated by degeneracies with the uncertain baryonic physics governing star formation.
In this work, we use a flexible astrophysical model and a Bayesian inference framework to analyse ultra-violet (UV) luminosity functions (LFs) at $z=6-8$. We vary the complexity of the galaxy model (single vs double power law for the stellar -- halo mass relation) as well as the matter power spectrum (cold DM vs thermal relic warm DM), comparing their Bayesian evidences.
Adopting a conservatively wide prior range for the WDM particle mass, we show that the UV LFs at $z=6-8$ only weakly favour CDM over WDM.  We find that particle masses of $\lesssim2$~keV are rejected at a 95\% credible level in all models that have a WDM-like power spectrum cutoff.  This bound should increase to $\sim2.5$~keV with the {\it James Webb Space Telescope (JWST)}.

\end{abstract}

% Select between one and six entries from the list of approved keywords.
% Don't make up new ones.

%%%%%%%%%%%%%%%%%%%%%%%%%%%%%%%%%%%%%%%%%%%%%%%%%%

%%%%%%%%%%%%%%%%% BODY OF PAPER %%%%%%%%%%%%%%%%%%

\section{Introduction}

The existence of dark matter (DM) has been proven by an enormous set of astrophysical and cosmological observational data. In the standard cosmological $\Lambda$CDM paradigm, dark matter particles are considered massive, collisionless and non-relativistic. In this scenario, cosmic structures are formed hierarchically from initial density fluctuations by the merging of smaller objects into larger ones. $\Lambda$CDM provides a good description of Cosmic Microwave Background (CMB) and large-scale structure (LSS) observations.  

Nevertheless, some physically-motivated extensions of the Standard model provide suitable DM candidates with masses in keV range, such as sterile neutrinos \citep[see, e.g.,  reviews in][]{Adhikari:17, Boyarsky:2018a} or gravitinos \citep{Viel:05}. These particles are initially relativistic and subsequently become non-relativistic before the matter-dominated epoch. The matter power spectrum of this so-called warm dark matter (WDM) is strongly suppressed on scales below the `free-streaming' length. Thereby WDM predicts a dearth of small-scale structures compared to CDM, potentially alleviating some (putative) tensions between CDM and observations on small scales (see, e.g., \citealt{Bullock:17, Kim:18}). Another motivation for  WDM candidates is the tentative detection of the $3.5$~keV line in the X-ray spectra of DM dominated objects \citep{Boyarsky:2014jta, Bulbul:2014sua}, which may be explained by the decay of $7$~keV sterile neutrinos.

If warm dark matter is in the form of thermal relics (e.g. gravitinos), the matter power spectrum is connected to that of CDM via a transfer function, which depends on the WDM particle mass.  As the particle mass increases, the transfer function cut-off scale moves to higher $k$ and predictions in WDM cosmologies become closer to those of CDM. On the other hand, sterile neutrinos do not reach thermal equilibrium with other particles, and their power spectrum also depends on the production mechanism.  However, the impact on structure formation is still through an effective suppression of the matter power on small scales.

Constraints on WDM properties generally make use of observations of non-linear small-scale structure, including the Lyman-$\alpha$ forest \citep{Viel:05, Viel:13, Baur:16, Irsic:17, Baur:2017stq, Garzilli:19}, Milky Way stellar streams \citep{Banik:19}, Milky Way satellites count \citep{Jethwa:16,Nadler:20,Newton:20, Nadler:21}, number counts and luminosity functions of distant galaxies \citep{Pacucci:13, Schultz:14, Menci:16a, Menci:16b, Corasaniti:17, Menci:17}, gamma-ray bursts \citep{deSouza:13}, strong gravitational lensing \citep{Gilman:19a,Gilman:19b}; see e.g. \cite{Enzi:20} for a recent review of current constraints.  
Hierarchical structure formation implies that "typical" structures are smaller (less massive) at higher redshifts.  This makes high-redshift observations particularly appealing in constraining WDM. 

Unfortunately, in all cases, degeneracy with astrophysics makes robust WDM constraints very challenging (though see the model-independent limits introduced in \citealt{Pacucci:13}). For example, the Lyman-$\alpha$ flux power spectrum depends on both the DM particle mass as well as the unknown thermal history of the IGM.  Similarly, galaxy luminosity functions depend on both the abundance of DM halos as well as the unknown star formation processes inside them (e.g. \citealt{Dayal:15, Corasaniti:17, Villanueva-Domingo:17, Khimey:20}).

In this work we re-visit WDM constraints implied by high-$z$ UV LFs, within a Bayesian framework  \citep[for a review of Bayesian methods in cosmology see, e.g.][]{Trotta:17}.  In addition to deriving lower bounds on the WDM particle mass\footnote{Throughout this paper, we assume WDM in the form of thermal relics for simplicity. Nevertheless, this approach can be applied for other dark matter models that predict the suppression of small-scale structures, e.g. ultra-light dark matter \citep[see, e.g.,][]{Hu:00,Marsh:14, Marsh:16, Schive:16}, interacting dark matter, etc. \citep[see, e.g.,][]{Bohm:02, Wilkinson:14, Schaeffer:21}. Furthermore, despite the fact that the ultra-light fuzzy dark matter transfer function has a different shape from that of thermal relic WDM  \citep[see, e.g.][]{Hu:00}, constraints on the WDM mass may be converted to those of the fuzzy DM particle mass, \citep[see, e.g.][]{Marsh:16,Armengaud:17}.
}, we also use the Bayesian evidence as an Occam's razor when performing model comparison in WDM and CDM.  The Bayesian evidence naturally penalises needlessly complicated models.
We vary the complexity of both the matter power spectrum parameterization (i.e. DM model) as well as the galaxy formation model, using the Bayes factor to penalise the "fine-tuning" provided by the addition of unnecessary parameters to the models.

This paper is organised as follows: in Sec.~\ref{sec:method} we describe our astrophysical model of the UV LFs, observational data-sets and Bayesian framework. In Sec.~\ref{sec:results} we report the results of the model comparison and WDM particle limits. We compare the obtained limits with those reported in other works in Sec.~\ref{sec:comparelimits}. Finally, we conclude in Sec.~\ref{sec:conclusions}.
Throughout this paper we assume \textit{Planck-16} cosmological parameters: $\Omega_{\Lambda}=0.685$, $\Omega_m=0.315$, $\Omega_b=0.049$, $h=0.673$, $n_s=0.965$, and $\sigma_8=0.83$ \citep{Planck:2015xua}. 

\section{Methodology}\label{sec:method}

\subsection{Modelling the ultra-violet luminosity functions}

To model the UV LFs, we use the simple galaxy model of \citet{Park:18} that ties galaxy properties to host halo masses via empirical scaling relations.  Specifically, we assume that the typical stellar ($M_\star$) to halo mass ($M_\text{h}$) relation of faint galaxies is characterized by a single power law (PL):
\eq{M_\star=f_\star\left(\frac{\Omega_\mathrm b}{\Omega_\mathrm m}\right)M_\mathrm{h},}
with
\eq{f_{\star}(M_\mathrm{h})=f_{\star10}\left(\frac{M_\mathrm{h}}{10^{10}M_\odot}\right)^\alpha.}

The typical star formation rate (SFR) of a galaxy hosted by a halo with mass $M_\text{h}$ is then expressed as
\begin{equation}
\label{eq:SFR}
\dot{M}_\star=\frac{M_\star}{tH^{-1}(z)} ~ ,
\end{equation}
where $t$ is dimensionless parameter which lies between 0 and 1, and $H(z)$ is the Hubble parameter.  The SFR and UV luminosity are related by $\dot{M}_\star=\kappa_\text{UV}L_\text{UV}$, where $\kappa_\text{UV}=1.15\times10^{-28}M_\odot$yr$^{-1}$ergs$^{-1}$s$^{-1}$Hz$^{-1}$ is a constant determined by the stellar initial mass function (e.g. \citealt{Sun:16}).
The absolute magnitude $M_\text{UV}$ is related to the UV luminosity with $M_\text{UV}=20.65-2.5\mathrm{log}_{10} \frac{L_\text{UV}}{\text{Hz}^{-1}\text{erg}\text{s}^{-1}}$.

Given the above, the UV LF can be constructed from the halo mass function (HMF; $\mathrm{d}n/\mathrm{d}M_\text{h}$) with:
\eq{\phi_\text{UV}=f_\text{duty} \frac{\mathrm{d}n}{\mathrm{d}M_\text{h}} \frac{\mathrm{d}M_\text{h}}{\mathrm{d}M_\text{UV}},}
where $f_\text{duty}$ describes the suppression of star formation in halos smaller than some characteristic scale $M_{\rm t}$ set by inefficient gas cooling, photo-heating of gas and/or SNe feedback \citep[see, e.g.][]{Okamoto:08, Sobacchi:13a, Sobacchi:13b, Dayal:14, Yue:16, Ocvirk:20}:
\eq{f_\text{duty}=\mathrm{exp}\left(-\frac{M_\text{t}}{M_\text{h}}\right).\label{eq:duty}}
Here, $M_\text{t}$ is the characteristic mass scale of the suppression.  

Because the normalisation of the stellar-to-halo mass relation and the characteristic star formation timescale are degenerate when computing the SFR (c.f. eq. \ref{eq:SFR}), we will define their ratio as $r_s \equiv f_{\star, 10} / t_\star$.  Therefore this simple UV LFs model has only three free parameters:
\begin{itemize}
\item $r_s \equiv f_{\star, 10} / t_\star$ - the ratio of the stellar fraction in $10^{10} M_\odot$ halos and the characteristic star formation time-scale (normalised by the Hubble time),
\item $\alpha_\star$ - the power law index of the stellar-to-halo mass relation,
\item $M_\text{t}$ - the characteristic turnover halo mass scale below which star formation is exponentially suppressed.
\end{itemize}
Although simple, this three parameter model can reproduce current high-$z$ LF observations \citep{Park:18, Oesch:18,Bouwens:2021}. Below we use this as our default galaxy model, referring to it as the {\it single power law model (PL)}. Both semi-analytic models and hydrodynamic simulations show that similar, simple power-law scaling relations capture the average properties of the high-redshift, faint galaxies of interest to us \citep[see][]{Behroozi:13, Mutch:16,Moster:13, Sun:16, Tacchella:18, Behroozi:19, Yung:19b}.

\begin{figure}
    \centering
        \includegraphics[width=.99\columnwidth]{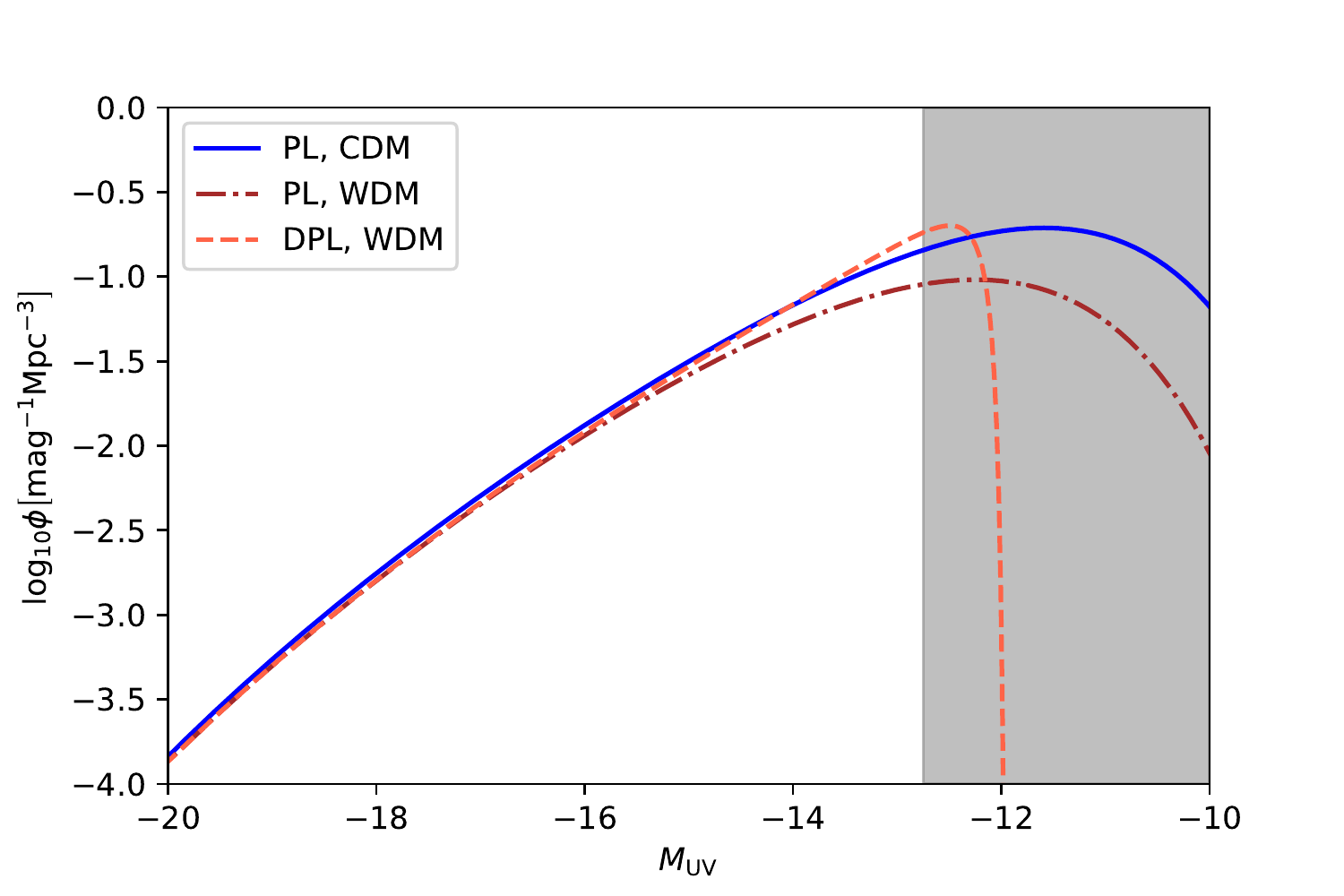}

    \caption{Examples of UV LFs from our models at z=6.  All cases assume inefficient star formation below a characteristic mass scale of $M_\mathrm{t} = 10^9~M_\odot$. The solid line corresponds to a our fiducial single-power law model (PL) for the stellar to halo mass relation and a CDM cosmology.  The dot-dashed and dashed lines correspond to WDM with $m_x = 3$ keV; 
    however, the latter uses a more complicated double power law (DPL) parametrization for the stellar to halo mass relation.
    The shaded region corresponds to the magnitudes which are not probed by HST observations.
     This example illustrates that an increase in star formation efficiency for faint galaxies can partially compensate for the decrease in galaxy number in WDM cosmologies over the observable range; however, the Bayesian evidence could penalise the red curve compared to the blue, due to its larger effective prior volume.}
    \label{fig:sketch_fs_DM}
\end{figure}

To check the robustness of our analysis, apart from the single power-law, we also consider a more complicated, {\it double-power-law (DPL)} star formation efficiency
\begin{equation}
f_\star=\frac{f_{\star\text{c}}}{2}\left[\left(\frac{M_\text{h}}{M_\text{c}}\right)^{\alpha_1}+\left(\frac{M_\text{h}}{M_\text{c}}\right)^{\alpha_2}\right].
\label{eq:dpl}
\end{equation}
This form allows one to increase the star formation rate in low-mass galaxies, such as might be the case if the EDGES signal \citep{Bowman:18} is proven to be of cosmic origin \citep{Mirocha:19, Quin:20a}. %(Mirocha and Furlanetto 2018; Qin et al. 2020).  
However, the main purpose for this study is that the additional flexibility of the DPL allows an increased star formation efficiency to partially compensate for a dearth in the number of DM halos in the WDM scenario.  An example of this is shown in Fig. \ref{fig:sketch_fs_DM}. Here the DPL WDM model denoted by the dashed curve mimics the (simpler) PL CDM model shown by the blue curve over the magnitude range probed by current observations.  The increased luminosity of the faint galaxies in the DPL model effectively moves them over to the left in this plot, resulting in the evident up-turn in the LF at $M_{\rm UV}>-15$ and a sharp cutoff beyond $M_{\rm UV}>-12$.

Even for CDM, we do not know the shape of the inevitable turn-over in the star formation efficiency at the faint end, and the DPL allows for additional flexibility; however, this flexibility comes at the cost of two additional parameters: a second power law index and the transition scale, $M_\mathrm{c}$.\footnote{Note that our DPL differs from the  double power law scaling used to characterize a turn-over in the star formation efficiency in massive galaxies, commonly attributed to AGN feedback (e.g. \citealt{Moster:13,Mirocha:16, Tacchella:18}).  Here we are only interested in the faint end of the LF ($M_{\rm UV}$> -20), where the putative supression of halos in WDM cosmologies would be most pronounced.}

In Appendix~\ref{sec:z-dep} we also show results using a redshift dependent stellar-to-halo mass relation.  Consistent with previous work \citep{Park:18, Oesch:18,Bouwens:2021}, we find that current data does not favour a redshift dependence of $f_\ast$.

\subsection{Halo mass function}

The halo mass function is defined as
\eq{\frac{\text{d}n}{\text{d}\text{ln}M_\text{h}}=f(\nu)  \frac{\rho_{m}}{M_\text{h}}\frac{\mathrm{d}\mathrm{ln}\sigma^{-1}}{\mathrm{d}\mathrm{ ln} M_\text{h}},}
where  $\nu=\left(\frac{\delta_c^2(z)}{\sigma^2}\right)$, $\delta_c(z)=\frac{1.686}{D(z)}$, $D(z)$ is the growth factor \citep{Heath:1977}, $\sigma^2(M)$ is the mass variance on scale $M$ and $\bar{\rho}_{m}$ is the mean matter density of the Universe.
In WDM cosmologies, the halo mass function cannot be analytically derived from the first principles. However, $N$-body siumulations show good agreement with analytic HMFs using modified transfer functions, over the range probed by current LF observations (e.g. \citealt{Schneider:14}).  Motivated by these results, we use the ellipsoidal collapse parametrisation (\citealt{Sheth:01})%; see \citealt{Schneider:14} for applications to WDM):
\eq{f(\nu)=A_\text{ST}\sqrt{\frac{2q\nu}{\upi}}\left(1+(q\nu)^{-p}\right) \mathrm{e}^{-q\nu/2},}
with $A_\text{ST}=0.322$, $p=0.3$, $q=1$ and $\nu=\frac{\delta^2_\text{c}(z)}{\sigma^2(M)}$.\footnote{We  confirm that our conclusions are unchanged using alternate values of the parameters, 
see~Appendix~\ref{sec:alt_hmf}.} 
To calculate $\sigma$, we use a sharp k-space filter, which is in a good agreement with the results of $N$-body simulations, over the relevant halo mass range \citep{Benson:12,Schneider:14, Bose:16}:
\eq{\sigma^2(M)=\int{P(k)\theta(1-kR)\mathrm{d}^3k},}
where  $M_\text{h}(R)=\frac{4\upi}{3}\left(\frac{R}{a}\right)^3$ and $\theta(x)$ is the Heaviside step function. Following \cite{Schneider:14}, we choose $a=2.5$.

The warm dark matter power spectrum is connected to the CDM one via the transfer function $T(k)$: 
\begin{equation}
P_{\text{WDM}}(k)=P_\text{CDM}(k)T^2(k).
\end{equation}
We use the transfer function parametrisation from \cite{Viel:05}:
\eq{T^2(k)=(1+(b k)^{2\mu})^{-10/\mu},}
where
\eq{b=0.049\left(\frac{m_\text{x}}{1\text{kev}}\right)^{-1.11}\left(\frac{\Omega_{\text{WDM}}}{0.25}\right)^{0.11}\left(\frac{h}{0.7}\right)^{1.22}h^{-1}\text{Mpc},}
$\mu=1.12$, and $m_\text{x}$ is the mass of the dark matter particle (in keV).\footnote{Strictly speaking, CDM also has a cut-off scale, which corresponds to a much shorter streaming length than WDM. However, this scale is far too small to impact the observations under consideration. Since our cosmological model is tied to the observations via the matter power spectrum, and WDM requires adjusting an additional turn-over scale to match the data, there is an Occam's razor penalty in the Bayesian evidence for WDM compared to CDM.}
We use \textsc{hmf} package for generation of halo-mass functions \citep{Murray:13}

\subsection{Data analysis}

We perform Markov chain Monte-Carlo (MCMC) sampling using the affine-invariant ensemble sampler implemented in the \textsc{emcee} code \citep{Foreman-Mackey:12}.
We use a likelihood function of the following form:
\begin{equation}\label{eq:likelihood}
    P(D|\theta,\mathcal{M})=\prod_{M_\text{UV}~\text{bins}}S\left(y_\text{pred}(\theta,\mathcal{M}),y_\text{obs},\sigma_1,\sigma_2,\right),
\end{equation}
where $y_\text{pred}(\theta,\mathcal{M})=\mathrm{log}_{10}\phi_\text{pred}(\theta,M)$ is the predicted value of the decimal logarithm of the UV LF in the model $\mathcal{M}$, $y_\text{obs}$  is the observational value, $\sigma_1$ and $\sigma_2$ are positive and negative errors. To account for asymmetric error bars, we take $S\left(y(\theta), \mu, \sigma_1, \sigma_2\right)$ to have a split-norm distribution (c.f. \citealt{Gillet:19}):

\begin{equation}\label{eq:likelihood-form}
    S\left(x,\bar{x}_\text{obs},\sigma_1, \sigma_2\right)=
    \begin{cases}
    A\mathrm{exp}\left(-\frac{(x-\bar{x}_\text{obs})^2}{2\sigma_1^2}\right), & x>\mu,\\
    A\mathrm{exp}\left(-\frac{(x-\bar{x}_\text{obs})^2}{2\sigma_2^2}\right), & x\leq\mu.
    
    \end{cases}
\end{equation}
Here $A=\frac{1}{\sqrt{2\upi}}\frac{2}{\sigma_1+\sigma_2}$ is the normalisation constant.

If we assume a single power-law $f_\star$ dependency on halo mass (model PL), the astrophysical model includes 3 free parameters: $r_\text{s}$, $\alpha$, $M_\mathrm{t}$. 
In case of DPL, the UV LF model includes 5 parameters, namely $r_\text{s},~\alpha_1,~\alpha_2,~M_{\text{c}},M_{\text{t}}$. In thermal relic WDM cosmologies, the matter power spectrum has an additional cutoff scale parametrized by the WDM mass $m_\text{x}$. As there is no reliable upper limit on the WDM mass, we adopt the inverse quantity $1/m_\text{x}$ as the free parameter\footnote{We also repeat our analysis with a flat prior over the range $m_\text{x}=$ 1--5 keV, as well as $m_\text{x}=$ 1--10 keV. We find that our model comparison results and derived 95\% $m_\text{x}$ constraints are unchanged by these alternate choices of prior.}. The upper limit on this parameter arises from the lowest allowed value of $m_\text{x}$, which we choose to be $m_\text{x}=1$~keV following numerous studies of structure formation~\citep[see e.g.][]{Enzi:20}. Our assumptions about the ranges of the values of model parameters and prior probability distributions are listed in Table~\ref{tab:1}.

\begin{table}
\centering
\begin{tabular}{c|c|c|c}
  Parameter   & Allowed range & Units & Prior \\
  \hline
   $r_\text{s}$& \,$0.01 - 100$ &  --  & flat log \\
   $\alpha$, $\alpha_1$, $\alpha_2$ & $-1.0 - 1.0$ &  -- & flat linear\\
   $M_\mathrm{t}$, $M_\mathrm{c}$ & $10^8 - 10^{10}$ & $\mathrm{M}_\odot$ & flat log \\
   %$M_\mathrm{c}$ & $10^8$ -- $10^{10}$ & $\mathrm{M}_\odot$ & flat log \\
     $1/m_\text{x}$   & $ 0 - 1 $ & keV$^{-1}$ & flat linear \\
%     $\gamma$ & $-3/2$ -- $3/2$ & - & flat linear \\
       \hline

\end{tabular}
\caption{The parameters of the models, used in this paper, and corresponding prior distributions.}
\label{tab:1}
\end{table}

\begin{figure}
    \centering
    \includegraphics[width = \columnwidth]{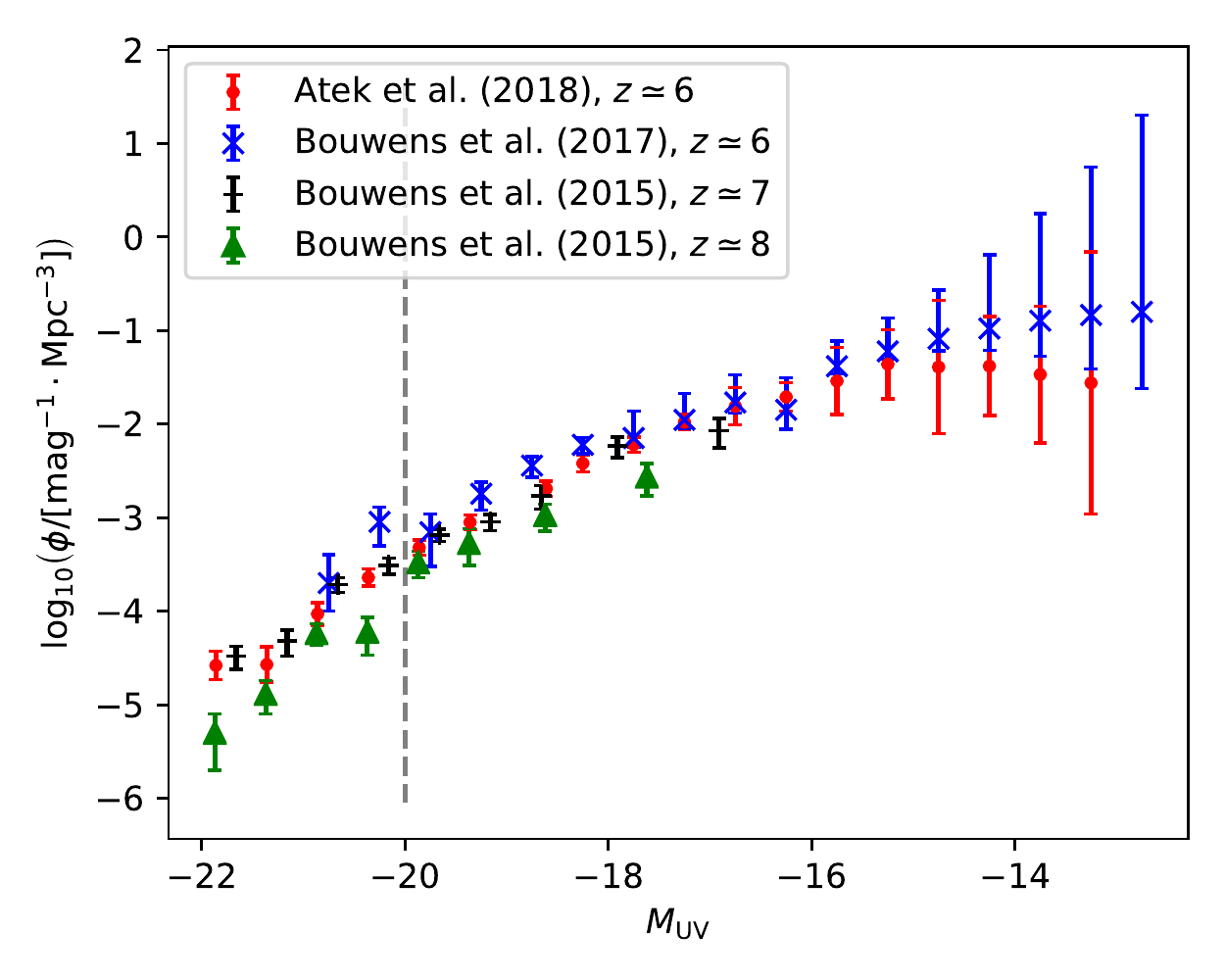}

    \caption{'A18' and 'B+' datasets. At $z=6$, the 'B17' and 'B+' datasets are the same. The dashed line shows the cut at magnitude $-20$ (see main text).}
    \label{fig:data}
\end{figure}

We analyse the following datasets:
\begin{itemize}
    \item 'A18' dataset is ultra-violet luminosity function  from  \cite{Atek:18} for $z\approx6$;
    \item 'B17' dataset is UV LF data from  \cite{Bouwens:17} for $z\approx6$
    \item 'B+' is 'B17' dataset at $z\approx6$ plus data from \cite{Bouwens:15} for $z=7$,~$8$.
    \item 'combined' corresponding to a concatenation of 'A18' and 'B+'.
\end{itemize} 
The corresponding data-points are shown in Fig.~\ref{fig:data}. Note that, by adding higher redshifts in the $B+$ dataset, we can probe also the redshift evolution of the LFs, which could better discriminate WDM and CDM \citep{Dayal:15}.  The 'combined' dataset equally weighs all observational data points in all of the data sets (see e.g. \citealt{Finkelstein:16} for a similar approach).

Our model of UV LF does not take into account the effects of AGN feedback nor dust extinction. Therefore we do not include data points with magnitudes lower than $-20$ focusing instead on the faint end that is most sensitive to WDM (c.f. Fig. 1). 
 Indeed, both observations and simulations suggest that dist attenuation is negligible at $M_\text{UV}>-20$ for the high redshifts of interest here (e.g. \citealt{Bouwens:14, Wilkins:17, Ma:19, Vijayan:21}).  Similarly, the impact of AGN feedback should be small on these magnitudes \citep[see, e.g.,][]{Yung:19a}.

We use the Bayes factor $K_{21}=\frac{P(M_2|D)}{P(M_1|D)}$ for model comparison. 
According to the Bayes theorem, the posterior probability of the model $M_i$ is defined as:
\begin{equation}
    P(M_i|D)=\frac{P(D|M_i)P(M_i)}{P(D)},
\end{equation}
therefore the Bayes factor could be expressed as a ratio of marginalised likelihoods (evidences):
\begin{equation}
    K_{21}=\frac{P(D|M_2)}{P(D|M_1)}, 
\end{equation}
assuming equal model priors $P(M_1)=P(M_2)$.
The model $M_2$ can be considered to be substantially supported if $\mathrm{log}_{10}K_{21}>1/2$ and strongly supported if $\mathrm{log}_{10}K_{21}>1$~\citep{Kaas:95}.

For each of the models under consideration, we calculate the marginalized likelihood 

\begin{equation}
P(D|M_i)=\int p(D|\theta,M_i)\pi(\theta,M_i)\mathrm{d}\theta,
\end{equation}
where $p(D|\theta,M_i)$ is the likelihood function, $\pi(\theta,M_i)$ are prior distributions of the parameters of the model $M_i$. The calculations are done by using dynamical nested sampling method~\citep{Higson:19} implemented in the \textsc{dynesty} \textsc{Python} package~\citep{dynesty}.

\section{Results}\label{sec:results}

\subsection{Model comparison}

\begin{figure*}
    \centering
    \includegraphics[width=2\columnwidth]{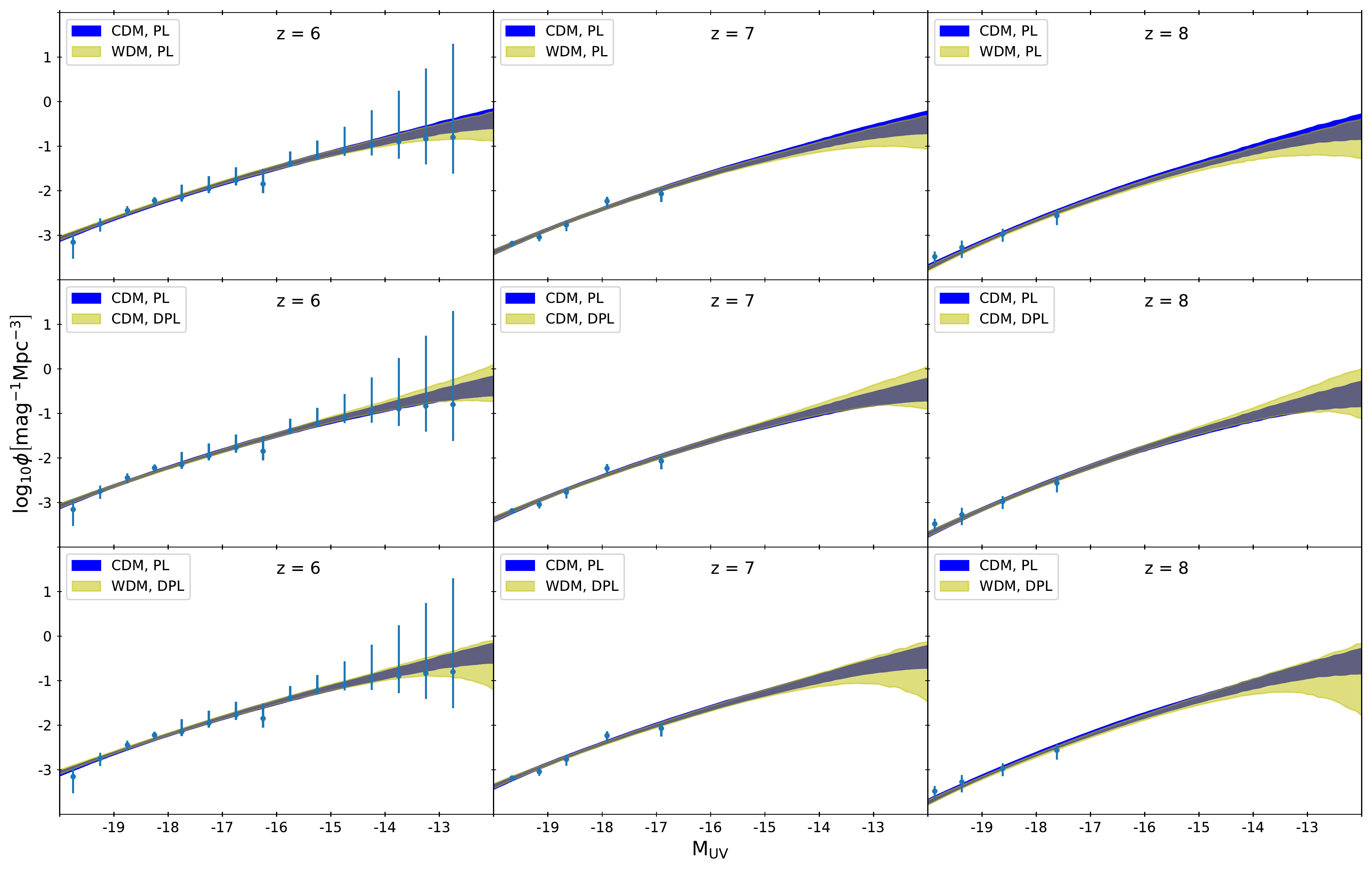}
    \caption{The 68\% C.L. of the UV LF posteriors for all models, computed using the `B+' data set. 
    Different redshifts are shown in columns, while rows correspond to different astrophysical and dark matter models.  For reference, the simplest PL + CDM model is shown in blue in all panels.  As expected, the posteriors of models with larger prior volumes (more free parameters) are broader, though the differences are only evident at $M_{\rm UV} < -15$.}
    \label{fig:uv_lf_ranges}
\end{figure*}

As an illustration of the procedure, in Fig.~\ref{fig:uv_lf_ranges} we show the 68\% C.L. of the marginalized LFs posteriors for each model (see Appendix~\ref{sec:posteriors} for the corresponding corner plots of the model parameters), using the 'B+' data set.  Columns correspond to different redshifts, while different astro/cosmo models are shown with the yellow shaded regions in the rows.  In all panels the simplest (CDM, PL) model is shown in blue.

The recovered LFs at the brighter end ($M_{\rm UV} < -15$) are comparable for all models considered.  However, the LF posterior is broader at $M_{\rm UV} > -15$ for more complicated models (larger prior volume).  This is especially evident in the bottom row, corresponding to the DPL + WDM model that has the most free parameters.  For example, at $M_{\rm UV} \gsim -13$ the DPL + WDM 68\% C.L. are wider by factors of $\gsim 2$.

\begin{table}
    \centering
    \begin{tabular}{lcccc}
            & \multicolumn{2}{c}{CDM}&\multicolumn{2}{c}{WDM}\\

          & PL & DPL & PL & DPL \\
         \hline
         
A18 &       0 &  $0.14\pm0.09$ &   $0.05\pm0.09$ &   $0.05\pm0.08$ \\

B17 &       0 &  $0.08\pm0.08$ &  $-0.33\pm0.09$ &  $-0.26\pm0.09$ \\
B+  &       0 &  $0.29\pm0.09$ &   $-0.33\pm0.1$ &   $-0.06\pm0.1$ \\

    \end{tabular}
    \caption{Base 10 logarithm of the Bayes factor for various dark matter scenarios and star formation efficiency models, with respect to the simplest (PL, CDM) model, $M_1$. The modest evidence ratios only allow for weak model preferences.}
    \label{tab:mlike}
\end{table}

How does their evidence compare?  We perform inference using all combinations of models and observational data sets. The resulting Bayes factors are summarised in Table~\ref{tab:mlike}, relative to the simplest CDM + PL model.  This table represents the main result of this work.

Unfortunately, we find that no Bayes factor is large enough to be used for conclusive model selection.
The CDM model is slightly preferred by the `B17' and `B+' datasets, but the corresponding Bayes factor $K\simeq 2$ is not enough to substantially support CDM against WDM.  Interestingly, the `B+' dataset substantially prefers CDM + DPL over WDM + PL; however, $\mathrm{log}_{10}\simeq0.6$ is very close to the lower level of Bayes factor required for substantial support of the model.

\subsection{Constraining the WDM particle mass}

\begin{figure}
     \centering
     \includegraphics[width=\columnwidth]{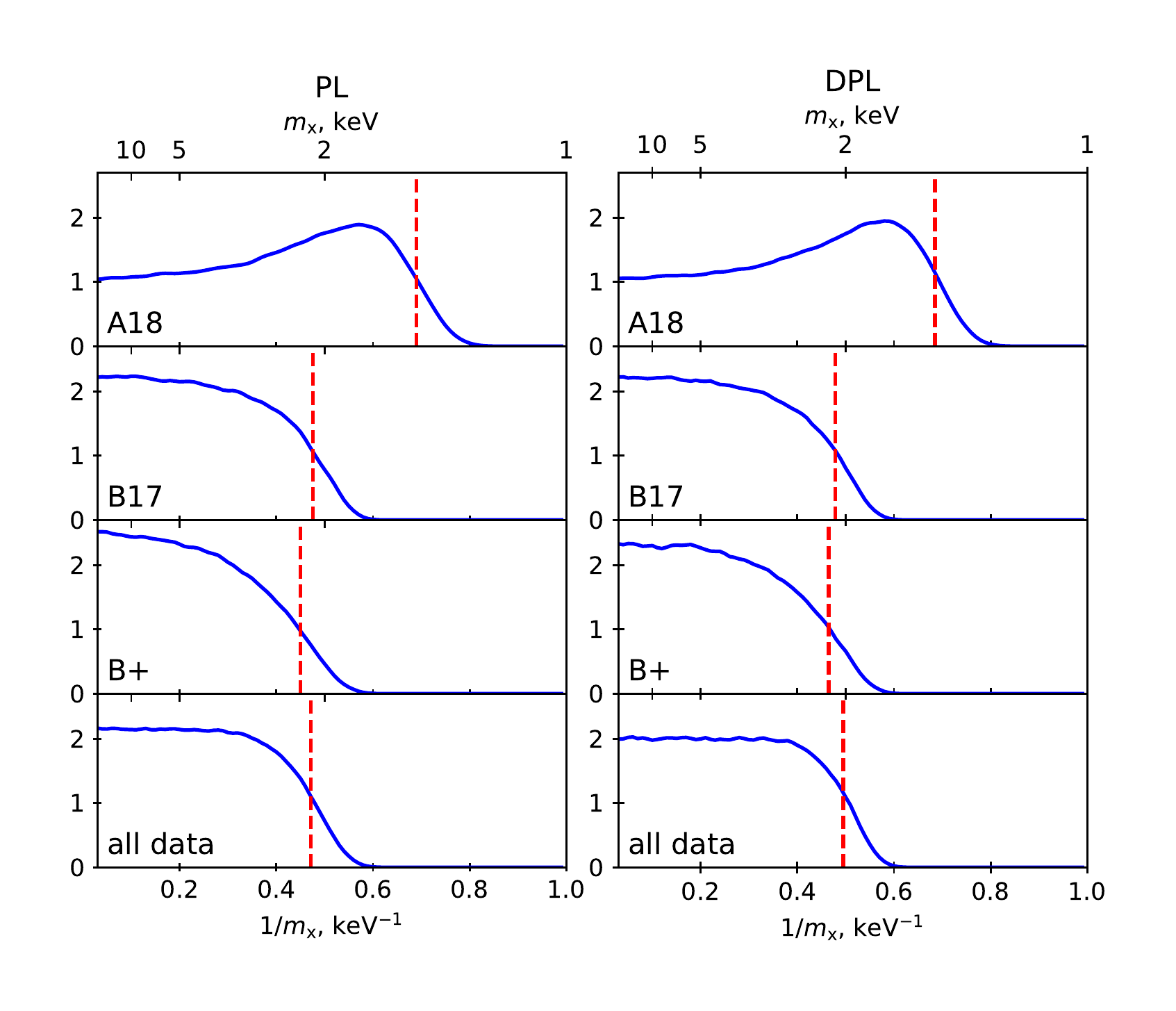}
    \caption{Marginalized posterior probability densities of $1/m_\text{x}$ for `power-law' and `double power-law' models of $f_\star$ and different datasets. Dashed lines denote the 95\% credible limits.}
    \label{fig:m_post}
\end{figure}

Given that we cannot rule out WDM using current LFs, we now look at the particle mass constraints in each of the WDM models.  The marginalized 1D posteriors of $1/m_{\rm x}$ for all WDM models and datasets are shown in Fig. \ref{fig:m_post}.  The 95\% C.L. are denoted by the vertical dashed lines are reported in Table \ref{tab:bounds}.
Due to its suggestive flattening of the faint-end slope (c.f. Fig. 2), the 'A18' dataset has the weakest limit: $m_\text{x}>1.5$\,keV.  All other datasets, including the combined one, result in $m_\text{x}\gsim2$\,keV.  Interestingly, the constraints using PL and DPL are comparable. This is likely do to the fact that the increased SFRs allowed by the DPL model can only modestly compensate for the dearth of halos in WDM; however, this occurs over the magnitude range where the observational error bars are very large (c.f. Figures 1 and 2).  For $m_\text{x}\lsim2$\,keV the dearth of halos is dramatic in the magnitude ranges probed by current observations, and cannot be compensated by astrophysics even in the DPL parameterization.

Understandably, we also find a strong degeneracy between $1/m_{\rm x}$ and the astrophysical turn-over scale $M_\mathrm{t}$ (see Figures B1 and B2). This is qualitatively consistent with previous works \citep{Dayal:17,Villanueva-Domingo:17, Esmerian:19, Khimey:20} that found a strong degeneracy between astrophysics and cosmology using currently available LFs. The  posteriors demonstrate that the constraints on both the WDM particle mass and turn-over mass mostly come from $z=6$ data, which goes to the faintest magnitudes. The addition of the higher redshift data mostly tightens constraints on 
 parameters responsible for the bright end of the LFs.

\begin{table}
    \centering
    \begin{tabular}{l|cccc}
     & `A18' & `B17' & `B+' & `combined' \\
    \hline
    PL & 1.5 & 2.1 & 2.2 & 2.1 \\
    DPL & 1.5 & 2.1 & 2.1 & 2.0 \\
    \end{tabular}
    
    \caption{95\% lower bounds (in keV) on the mass of thermal relic WDM particle resulting from different datasets and model choices.}
    \label{tab:bounds}
\end{table}

\subsection{How will limits improve with JWST?}

Future observations with the {\it James Webb Space Telescope (JWST)} should probe fainter magnitudes than currently available with \textit{HST}. This should facilitate tighter DM constraints (e.g. \citealt{Dayal:15, Lapi:15}).

To quantify this, we repeat our analysis using the CDM+PL and WDM+PL models on the simulated \textit{JWST} UV LF observations from \cite{Park:20} (specifically their \textit{JWST}-F dataset).  These authors use the results of hydrodynamic simulations that agree with current data, but provide predictions at magnitudes too faint to be currently observed with {\it HST}.  {\it JWST} faint end uncertainties were simply estimated by shifting the current HST error bars 1.5 magnitudes deeper (see \citealt{Park:20} and references therein for more details).

We find that this \textit{JWST} mock observation supports CDM over WDM with a Bayes factor of $\simeq3$, which is still not sufficient for a robust distinction between the models.  We also find that the 95\% C.L. lower bound on the WDM mass increases to  2.5~keV. Our results are in agreement with \cite{Khimey:20}.

\section{Comparison with existing WDM constraints}\label{sec:comparelimits}

It is useful to place our WDM particle mass limits in context with other limits. Several approaches have been applied to constrain WDM using UV LFs.  The most model-independent approach was suggested by \citet{Pacucci:13}: comparing the cumulative galaxy number to the cumulative halo number.  This approach is only dependent on the assumed HMF, and is conservative since it does not depend on how UV luminosity is assigned to DM halos.
\citet{Pacucci:13} initially quoted a limit of $m_{\rm x}>1$ keV based on preliminary results from the CLASH survey (see also \citealt{Schultz:14,Menci:16a,Menci:16b,Menci:17}).  Subsequent work suggested a $2.4$~keV  $2\sigma$ lower bound \citep{Menci:16b} using the LFs from \cite{Livermore:17}. Our constraint is somewhat weaker than this because the faint-end of the \cite{Livermore:17} LFs are systematically higher than the data sets used in this work.   Alternatively, \citet{Corasaniti:17} modelled the UV LFs in WDM, fitting the data from~\cite{Atek:15,Bouwens:15,Bouwens:17}.
Their 1.5\ keV lower bound is somewhat weaker than ours, possibly due to the different assumed halo mass functions.

  Tighter $3-5$~keV 2$\sigma$ lower limits on the WDM mass were obtained from the analysis of small-scale structure in the Lyman-$\alpha$ forest ~\citep[see, e.g.,][]{Viel:13, Baur:16, Irsic:17}. These constraints require marginalizing over the thermal and ionization history of the IGM. More recent, conservative assumptions on these histories resulted in a weaker $\sim2$~keV (2$\sigma$) limit \cite{Garzilli:19}, which is the same as we find here using the UV LFs.
 
Even tighter $4-5$~keV 2$\sigma$ lower bounds are obtained from the analysis of flux ratios of the quadruply-imaged quasars \citep{Gilman:19a,Gilman:19b}. However, these lower bounds are strongly dependent on the choice of prior \citep[see Sec.4.2 in][]{Enzi:20}.
To overcome this difficulty,~\citet{Gilman:19b,Enzi:20,Nadler:21} used marginalised likelihoods ratios to obtain somewhat weaker lower limits: $3-4$~keV in \citep{Gilman:19b}; and $2.68$~keV from the joint analysis of Lyman-$\alpha$, MW satellites and gravitational lensing in \cite{Enzi:20}. The recent work of \citet{Nadler:21} rejects $\lesssim 7.4$~keV masses with marginalized likelihood ratio between such WDM and CDM $\lesssim 1/20$ (roughly corresponding to the 2$\sigma$ bound) from the joint analysis of MW satellites and strong gravitational lensing. It should be noted, that these constraints are strongly dependent on the sub-halo mass function.  There is some concern that previously proposed fits of sub-halo mass functions, used in these works, underestimate the number of satellites~\citep{Lovell:19}. Furthermore, recent MW satalite analysis by \citet{Newton:20} gives a  a weaker but less model-dependent lower limit of $\sim 2$~keV (when marginalizing over baryonic feedback) and a stronger but model-dependent lower limit of $3-4$~keV (when modelling the baryonic feedback).

\section{Conclusions}
\label{sec:conclusions}

The abundances and brightness of early galaxies depend both on the unknown dark matter properties and  the poorly-constrained physics of star formation.  The degeneracy between astrophysics and cosmology is always a bottle-neck when constraining DM models with a dearth of small-scale power, such as WDM.

In this paper, we apply Bayesian inference to the analysis of UV LFs at redshifts $6-8$, varying prescriptions for both the dark matter and star formation.  We adopt two flexible astrophysical models, using a single and a double power-law to characterize the stellar-to-halo mass relation. For each astrophysical model, we also assume either CDM or thermal relic WDM when computing halo abundances.

We compute the Bayes factor between the models and find no substantial preference for any model, using current UV LFs. CDM is only weakly favoured over. We also conclude that our fiducial model (CDM with a  simple redshift-independent power-law star-formation efficiency for faint galaxies) describes the existing UV LF data well.

We find that in WDM cosmologies, particle masses of $\lesssim2$~keV are rejected with a 95\% credible level using only the UV LFs.  This result is consistent with other astrophysical limits on the particle mass.  Using a mock {\it JWST} dataset from \citep{Park:20}, we forecast that this limit could improve to $\sim$2.5 keV with upcoming {\it JWST} observations.

Our work showcases how Bayesian model comparison can be applied to reionization-era observations, acting as an Occam's razor to discriminate against needlessly complicated astrophysics that can mimic cosmology.  In the future, we will apply this framework to upcoming 21-cm interferometric observations of the EoR and Cosmic Dawn.   Although galaxy astrophysics can be "tuned" to mimic a 21-cm WDM signal, such tuning is likely very ad-hoc (see for example Fig. 5 in \citealt{Sitwell14}).  Indeed, \citet{Munoz:20} and \citet{Jones21} use a simpler SFR prescription and Fisher forecasts to suggest that upcoming interferometers would be able to constrain warm and fuzzy dark matter, respectively. The enormous data-set provided by upcoming 21-cm observations with HERA and SKA will be very constraining, facilitating detailed model selection \citep{Binnie19, Quin:20b}.

\section*{Acknowledgements}
We thank R. Trotta, A. Pilipenko and M. Viel for helpful comments.  AR and DS acknowledge support from the National Academy of Sciences of Ukraine by its priority project No.0120U100935  `Fundamental properties of the matter in the relativistic collisions of nuclei and in the early Universe'. The work of AR was also partially supported by the ICTP through AF-06. AM acknowledges funding from the European Research
Council (ERC) under the European Union’s Horizon 2020
research and innovation programme (grant agreement No
638809 – AIDA). The results presented here
reflect the authors’ views; the ERC is not responsible for
their use. The calculations presented here were performed on the BITP computer cluster.

\section*{Data availability}
The  luminosity functions  underlying  this  article are publicly available. The code that supports the findings of this study will be shared on reasonable request to the corresponding author.

%%%%%%%%%%%%%%%%%%%% REFERENCES %%%%%%%%%%%%%%%%%%

% The best way to enter references is to use BibTeX:
\bibliographystyle{mnras}
\bibliography{refs}

%%%%%%%%%%%%%%%%%%%%%%%%%%%%%%%%%%%%%%%%%%%%%%%%%%

%%%%%%%%%%%%%%%%% APPENDICES %%%%%%%%%%%%%%%%%%%%%
\appendix
\section{Redshift-evolution of star formation efficiency}\label{sec:z-dep}

In this work the base-line parametrisation considers the free parameters to be redshift-independent. The redshift-dependence of SFR is through the characteristic time-scale, i.e. $\dot{M}_\star=\frac{M_\star}{tH^{-1}(z)}$, which leads to $\dot{M}_\star\sim (1+z)^{3/2}$ during the matter dominated epoch \citep[see the motivation of this form in][]{Park:18}. While this assumption successfully describes the observed UV LFs, one could consider also a more flexible redshift evolution (e.g. \citealt{Mirocha:20}). 

Thereby, we additionally test a model with  $f_\star=f_\star(z)$ in the following form:
\begin{equation}
f_\star=f_{\star,6}(M_\text{h})\left(\frac{1+z}{7}\right)^\gamma\,,
\label{eq:fz_dep}
\end{equation}
where $\gamma$ is an additional free parameter, $f_{\star,6}(M_\text{h})$ is the star formation efficiency at $z=6$ with `PL' or `DPL' halo mass dependency. We assume a flat prior for $\gamma$ over the range $(-3/2,3/2)$ .

We calculate Bayes factors for WDM and CDM using this redshift-dependent star formation efficiency, with respect to the fiducial CDM with a redshift-independent $f_\star$.  The results are shown in Table~\ref{tab:z_dep}.
\begin{table}
    \centering
    \begin{tabular}{cccc}  \multicolumn{2}{c}{CDM}&\multicolumn{2}{c}{WDM}\\

           PL & DPL & PL & DPL \\
         \hline
         $-0.48\pm0.09$&$-0.18\pm0.09$&$-0.56\pm0.1$&$-0.73\pm0.1$\\
         
    \end{tabular}
    \caption{Log$_{10}$ of Bayes factor of CDM and WDM dark matter scenarios with `PL' and `DPL' $f_\star$ halo mass dependency and on the assumption of redshift dependency according Eq.\ref{eq:fz_dep}. The evidences are calculated on the B+ dataset. The CDM scenario with `PL' redshift-independent star formation efficiency is assumed as a fiducial model.}
    \label{tab:z_dep}
\end{table}

We don't find any support for such a redshift-dependent $f_\star$. Moreover, the WDM scenario with  a redshift-dependent $f_\star$ is substantially disfavoured. Also, the most probable values of $\gamma$ are close to 0 for all cases under consideration.  We thus conclude that there is currently no need for a redshift evolution in the star formation efficiency, consistent with previous works \citep{Park:18, Oesch:18,Bouwens:2021}.

\section{Halo mass function with alternative parametrisation}\label{sec:alt_hmf}
While the sharp $k$-space filter allows one to capture the low-mass end of WDM halo mass function reasonably well, the high-mass end  is systematically underestimated (e.g. up to $15-20$\% at $z=6$ using the Sheth-Tormen form; see more in \citealt{Schneider:13, Schneider:14}). 
We have found that the simulations from \cite{Schneider:14, Bose:16} are well-described on both high- and low-mass scales by the Sheth-Tormen approximation with the following parameters: $A_\textit{ST}=0.322$, $q=0.93$ and $p=0.3$. 

We repeat all of our calculations with these parameters, and do not find any significant difference with results obtained in our base-line model, see Table~\ref{tab:bounds_alternative_hmf},~\ref{tab:K_alternative_hmf}.
\begin{table}
    \centering
    \begin{tabular}{l|c|c}
         &WDM, `PL'  &WDM, `DPL'\\
         \hline
         A18 &  1.41 &  1.40 \\
         B17 &  2.03 &  2.03 \\
         B+  &  2.08 &  2.14 \\

    \end{tabular}
    \caption{The obtained 95\% lower bounds (in keV) on the WDM particle mass resulting from different datasets and model choices, using alternative halo-mass function parameters.}
    \label{tab:bounds_alternative_hmf}
\end{table}

\begin{table}
    \centering
    \begin{tabular}{l|c|c}
         &WDM, `PL'  &WDM, `DPL'\\
         \hline
        A18 &   0.13$\pm$0.09 &   0.01$\pm$0.09 \\
        B+  &  -0.14$\pm$0.09 &   -0.35$\pm$0.1 \\
        B17 &  -0.24$\pm$0.09 &  -0.18$\pm$0.09 \\

    \end{tabular}
    \caption{Base 10 logarithms of the Bayes factor, which is calculated for warm dark matter scenario with alternative halo-mass function parameters vs cold dark matter with `PL' $f_\star$. }
    \label{tab:K_alternative_hmf}
\end{table}

\section{Posterior probability distributions}\label{sec:posteriors}
\begin{figure*}
	\includegraphics[width=1.5\columnwidth]{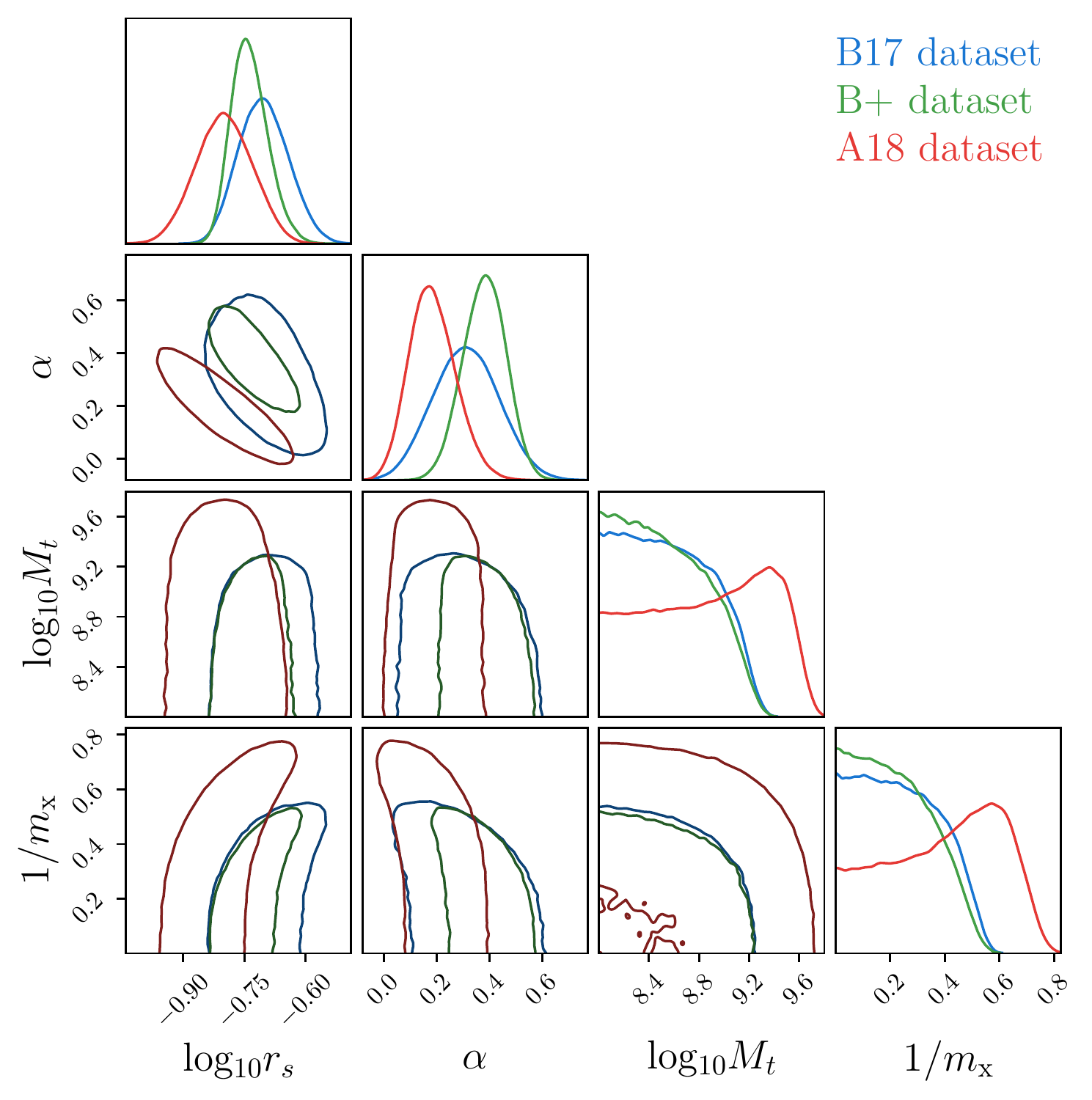}
    \caption{Corner plot showing parameter constraints, obtained from the MCMC analysis of A18, B17 and B+ datasets with the simple power-law star formation efficiency model (PL): the 95\% contours of the joint 2D marginalised posterior distributions with PDF along the diagonal.
}
    \label{fig:4_pars}
\end{figure*}

\begin{figure*}
	\includegraphics[width=2\columnwidth]{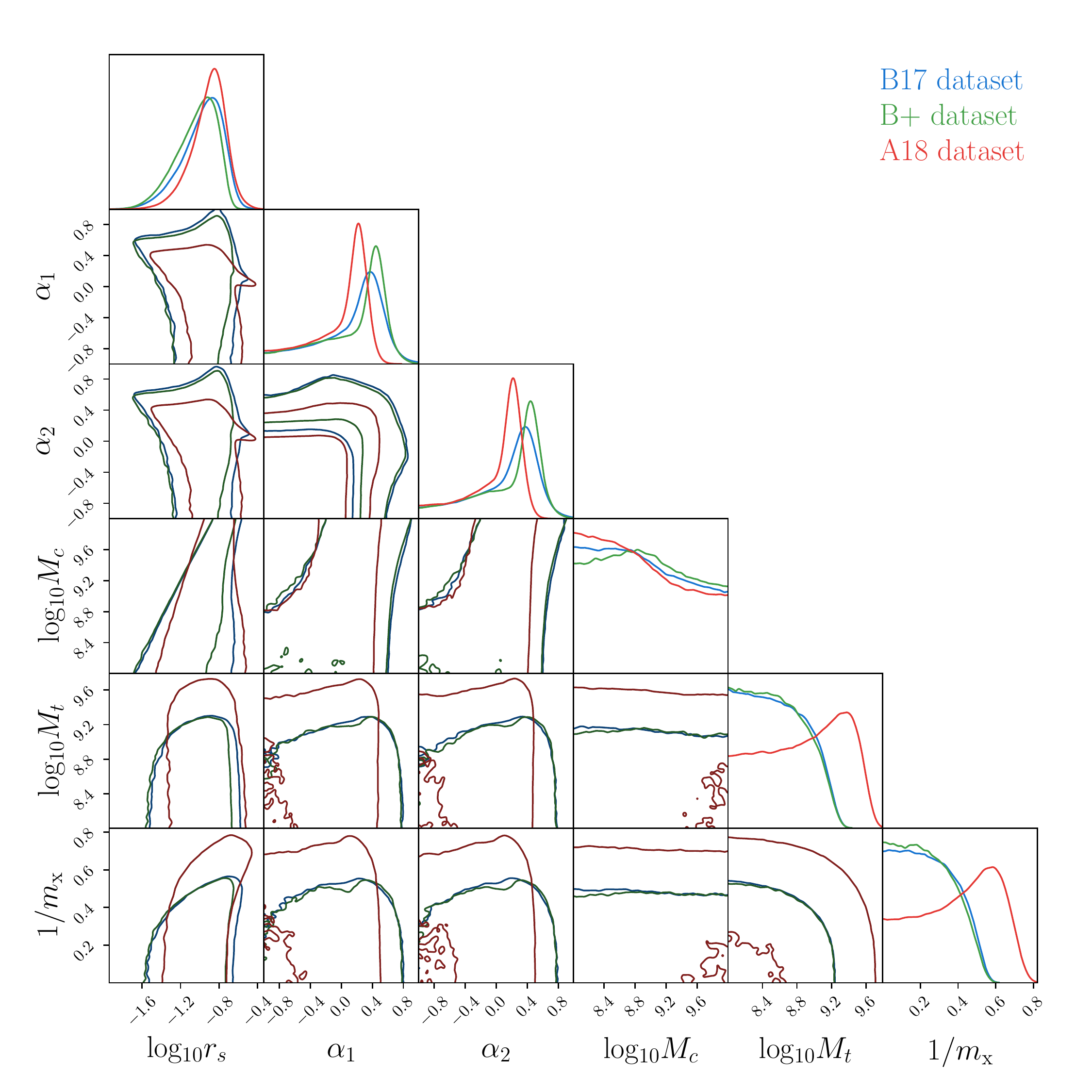}
    \caption{The same as in Figure~\ref{fig:4_pars}, but in the case of `double power-law' (DPL) model.
    }
    \label{fig:6_pars}
\end{figure*}

The corner plots corresponding to our astrophysical models, PL and DPL, and three datasets ('A18', 'B17', 'B+') are shown in Fig.~\ref{fig:4_pars} (PL $f_\star$) and Fig.~\ref{fig:6_pars} (DPL $f_\star$). 
The 2D marginalised posterior regions show the degeneracy between the DM free-streaming properties (encoded by the inverted WDM particle mass $1/m_\text{x}$) and the characteristic turnover halo mass scale set by baryonic physics, $M_\text{t}$. We confirm that the posteriors calculated via \textsc{dynesty} are fully consistent with the MCMC results, which additionally verifies the robustness of the inference.

%%%%%%%%%%%%%%%%%%%%%%%%%%%%%%%%%%%%%%%%%%%%%%%%%%

% Don't change these lines
\bsp	% typesetting comment
\label{lastpage}
\end{document}